\documentclass[aip,jcp,groupedaddress,a4paper,12pt,reprint]{revtex4-1}
\usepackage{amsmath}
\usepackage{amsfonts}
\usepackage{amssymb}
\usepackage{graphicx}

\usepackage{color}

\begin{document}
\title{First-principles thermodynamic screening approach to photo-catalytic water splitting with co-catalysts}
\author{Harald Oberhofer}
\email{harald.oberhofer@tum.de}

\author{Karsten Reuter}

\affiliation{Department Chemie, Technische Universit{\"a}t M{\"u}nchen, Lichtenbergstr. 4, D-85747 Garching, Germany}
\date{\today}

\begin{abstract}
We adapt the computational hydrogen electrode approach to explicitly account for photo-generated charges and use it to computationally screen for viable catalyst/co-catalyst combinations for photo-catalytic water splitting. The hole energy necessary to thermodynamically drive the reaction is employed as descriptor for the screening process. Using this protocol and hybrid-level density-functional theory we show that water oxidation on bare TiO$_2$ surfaces is thermodynamically more complex than previously thought. This motivates a screening for suitable co-catalysts for this half-reaction, which we carry out for Au particles down to the non-scalable size regime. We find that almost all small Au clusters studied are better suited for water photo-oxidation than an extended Au(111) surface or bare TiO$_2$ facets. \emph{Copyright (2013) American Institute of Physics. This article may be downloaded for personal use only. Any other use requires prior permission of the author and the American Institute of Physics. The following article has been accepted by the Journal of Chemical Physics. After it is published, it will be found at \href{http://jcp.aip.org/}{http://jcp.aip.org/}.}
\end{abstract}

\maketitle

\section{Introduction}
Oxide surfaces -- such as TiO$_2$ \cite{Fujishima1972NAT} -- are long known to possess the ability to split water using only light. Unfortunately, the yields are small and most of the reactive surfaces only work with highly energetic UV-light.\cite{Fujishima2008SSR,Henderson2011SSR} A common route to make the process more viable for large scale application is the introduction of suitable co-catalysts that offer reactive sites or act as carrier traps.\cite{Maeda2010JPCL} With usual noble metals (Pd, Pt, Au) already leading to a clear enhancement of the photo-catalytic activity,\cite{Kim2002JPCB,Sakthivel2004WR,Subramanian2004JACS} current research aims to identify alternative low-cost materials with equal or even better functionality. To this end the use of smaller and smaller nanoparticles \cite{Iwase2006CL,Hernandez2009EES} is not only appealing in terms of material efficiency, but also with respect to the intriguing nano-catalytic properties of metal clusters in the non-scalable size regime.\cite{Heiz2007BOOK} On the other hand the size dimension introduced in this regime further enhances the anyways huge chemical compound space of possible photo-catalyst/co-catalyst combinations. Recognizing that computational screening is an ever more powerful tool for such problems \cite{Norskov2009NATC} we here present an adaption of the computational hydrogen electrode approach of N\o{}rskov and Rossmeisl \cite{Norskov2004JPCB,Valdes08JPCC} in order to account for explicit photo-generated charges and efficiently screen for optimal co-catalyst/substrate combinations.

\section{Screening of photo-(co-)catalysts}
This approach is equally applicable to hydrogen evolution and water oxidation, but for reasons further specified below we here focus on the latter half-reaction. A generally considered pathway for this reaction proceeds via four electron-coupled proton transfer steps:
\begin{subequations}
	\label{eq:nr-orig}
	\begin{align}
		A: \quad & {\rm H_2O} + (^*) \rightarrow {\rm OH^*} + {\rm H^+} + e^- \label{eq:nr-stepa}\\
		B: \quad & {\rm OH^*} \rightarrow {\rm O^*} + {\rm H^+} + e^- \\
		C: \quad & {\rm H_2O} + {\rm O^*} \rightarrow {\rm OOH^*} + {\rm H^+} + e^- \\
		D: \quad & {\rm OOH^*} \rightarrow {\rm O_2} + (^*) + {\rm H^+} + e^- \quad ,
	\end{align}
\end{subequations}
where the asterisk stands for the catalytic surface $(^*)$ and particles attached to it (e.g.~O$^*$), respectively. At each of the four steps the respective adsorbate loses one proton to the surrounding medium, while one electron annihilates with a photo-generated hole localised on the substrate surface. Analogous to the original method by N\o{}rskov and Rossmeisl the idea of our density-functional theory (DFT) based approach is to determine the relative free energies and thus the relative stabilities of all reaction intermediates along this pre-defined pathway (eqs.~\ref{eq:nr-orig}). Given these free energy differences one can then -- disregarding any kinetic barriers -- predict if a reaction is energetically feasible or could get ``stuck'' in a particularly stable intermediate. The effect of the hole charges on the surface is thereby to ``drive'' the otherwise endothermic reaction at each step by lowering the overall free energy of intermediates and product. As such the approach not only allows to account for this effect, but also as the main result to predict the minimum hole energy $\epsilon_{\rm hole}$ necessary to make the reaction thermodynamically feasible, i.e. drive the reaction such that each step is downhill or at least level in free energy (see theory section below). In standard electrochemistry $\epsilon_{\rm hole}$ would relate to the overpotential, and in the present context we use it as a suitable descriptor to screen for viable co-catalyst/substrate combinations.

\subsection{Free energy expressions}
In the present work we consider photo-catalytic water splitting, i.e.~in the presence of hole carriers generated by photon absorption and subsequent exciton dissociation towards the respective electrodes. Additionally, we assume that charge diffusion in the semiconductor is fast compared to the oxidation reaction such that for each reaction step there is a hole present driving the reaction. Calculation of relative free energies of reaction intermediates is analogous to earlier work by N\o{}rskov, Rossmeisl, and co-workers:\cite{Norskov2004JPCB}
	
\begin{subequations}
	\label{eq:nr-Gfull}
	\begin{align}
		\Delta G_{A}=&(E_{\rm OH^*}-E_{^*})+\text{EA}_{s^+}-E_{\rm H_2O}+G_{\rm H^+}+S_{A}\\
		\Delta G_{B}=&(E_{\rm O^*}-E_{\rm OH^*})+\text{EA}_{s^+}+G_{\rm H^+}+S_{B}\\
		\Delta G_{C}=&(E_{\rm OOH^*}-E_{\rm O^*})+\text{EA}_{s^+}-E_{\rm H_2O}+G_{\rm H^+}+S_{C}\\
		\Delta G_{D}=& \Delta G_\text{\rm exp}+\text{EA}_{s^+}+(E_{^*}-E_{\rm OOH^*})+2E_{\rm H_2O}\nonumber\\
		&-2E_{\rm H_2}+G_{\rm H^+}+S_{D} \label{eq:dG} \quad,
	\end{align}
	\end{subequations}
where $E_{\rm X}$ are the DFT total energies of species X and $S_{Y}$ denote the total entropic and zero point energy contributions to reaction step $Y$. In eq.  ~\ref{eq:dG} the product ($O_2$) is expressed through the experimental water splitting free energy $\Delta G_\text{exp}$, analogous to N\o{}rskov and Rossmeisl in order to avoid the calculation of the -- in DFT only poorly described -- total energy of $\rm O_2$. The main difference between eqs.~\ref{eq:nr-Gfull} and the original electrochemical formulation is that instead of $1/2 E_{H_2}$ and an external potential, we use the solvation free energy of a proton ($G_{\rm H^+}=-11.53$\,eV) \cite{Cheng10PRB} plus the energy gain of an electron falling into a pre-generated hole on the catalytic surface ($\text{EA}_{s^+}$). The screening descriptor $\epsilon_{\rm hole}$ is then given by the value of $\text{EA}_{s^+}$ for which all $\Delta G_Y\leq 0$. In the present work we do not venture to calculate specific values of $\text{EA}_{s^+}$ for possible substrate surfaces, but rather focus -- in the spirit of materials screening -- on determining $\epsilon_{\rm hole}$ of a given cluster, in order to predict suitable cluster/substrate combinations.
Note also that the photo-catalytic surface only contributes through the energy of the surface-localised hole. This means that -- neglecting solvent and geometric effects as well as charge transfer between surface and co-catalysts -- it is sufficient to perform DFT calculations of only the co-catalyst/adsorbate complexes.
\subsection{TiO$_2$ revisited}

\begin{figure}[ht]
		\centering
		\includegraphics[scale=1.0,clip=true]{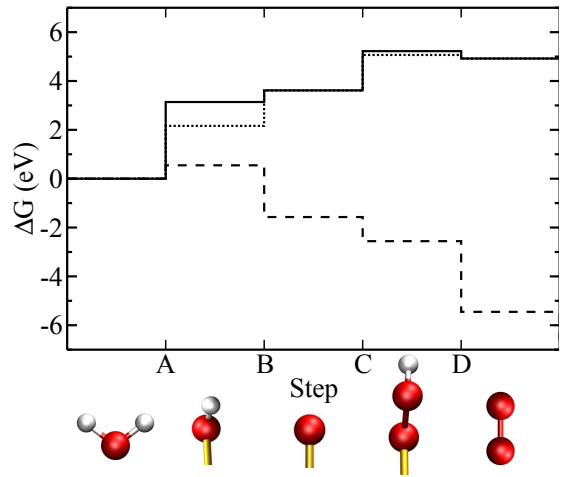}
		\caption{Calculated free energy changes along the water oxidation pathway at pH=0, cf. eqs.~(\ref{eq:nr-orig}), on the (110) 
		facet of rutile TiO$_2$. Solid line: Free energy profile without reaction-driving hole carriers calculated with PBE0, for comparison the free energy profile calculated with PBE is depicted as dotted line. Dashed line: Profile
		considering photo-generated holes at the valence band maximum of TiO$_2$ (taken as $\epsilon_{\rm VBM} = -7.1$\,eV).}
		\label{fig:dG-TiO2}
\end{figure}

Focusing first on the photo-catalyst we consider the prototypical (110) facet of rutile TiO$_2$ which has been studied in earlier work within the computational hydrogen electrode approach \cite{Valdes08JPCC}. Resulting free energy profiles with and without a hole driving the reaction at pH=0 are depicted in Fig.~\ref{fig:dG-TiO2}. Note that -- using hybrid-level DFT -- the initial step in the reaction sequence (eq. \ref{eq:nr-stepa}) is by far energetically dominant. In fact, this step $A$ is much more uphill than predicted in the GGA-based earlier work \cite{Valdes08JPCC} (cf.~also dotted line in Fig.~\ref{fig:dG-TiO2}), and therefore requires a hole energy of at least $\epsilon_{\rm hole} = -7.5$\,eV with respect to vacuum to drive the entire reaction downhill. This is an intriguing result if one considers that the experimental valence band maximum (VBM) of rutile TiO$_2$ is located at $\epsilon_{\rm VBM} = -7.1$\,eV ~\cite{Cheng10PRB}. Hybrid-level DFT would thus predict that photo-oxidation of water would not be thermodynamically feasible at defect-free rutile TiO$_2$(110). Notwithstanding, to put this into perspective one has to recognize though that due to the intrinsic problems of calculating and measuring accurate absolute band positions there is a considerable spread in literature concerning the TiO$_2$ VBM position \cite{Gratzel2001NAT, Cheng10PRB, Kang2010PRB}, with recent many-body perturbation theory based calculations even obtaining $\epsilon_{\rm VBM} = -8.0$\,eV~\cite{Kang2010PRB}. Remaining uncertainties in $\epsilon_{\rm hole}$ at hybrid-level DFT (e.g.~due to persistent charge delocalisation) certainly also add to the picture. On the other hand, the bare VBM of bulk TiO$_2$ is in fact not the fully appropriate reference as it neglects the necessity to localize the hole at the given surface. Zawadzki {\em et al.} have recently shown that the in any case pronounced hole self-trapping in TiO$_2$ is furthermore highly surface sensitive \cite{Zawadzki12EES}. The appropriate reference is thus the corresponding surface trapped level $\epsilon_{\rm T}$ in the TiO$_2$ band gap, as this will be the one predominantly populated by photo-generated holes. In the case of the rutile (110) surface the $\epsilon_{\rm T}$ computed by Zawadzki and co-workers lies only 0.2\,eV higher than $\epsilon_{\rm VBM}$, but at other commonly occurring rutile and anatase TiO$_2$ facets this is much larger \cite{Zawadzki12EES}. We correspondingly compute the water oxidation pathway also for these facets and contrast the results with the different hole trapping levels in Table \ref{tab:TiO2-hole}. Even though the $\epsilon_{\rm hole}$ are more approximate for these facets, as they have not specifically been computed for the coverages of relevance here \cite{Valdes08JPCC}, the result is quite striking. In all cases the available photo-generated holes would not thermodynamically be able to drive the water oxidation when using $\epsilon_{\rm VBM} = -7.1$\,eV ~\cite{Cheng10PRB} as value for the TiO$_2$ VBM. While a more negative VBM reference value, e.g. the $\epsilon_{\rm VBM} = -8.0$\,eV from Kang and Hybertsen \cite{Kang2010PRB}, would change this for some facets, it would only just be so, i.e. the available $\epsilon_{\rm T}$ would at best straddle the required $\epsilon_{\rm hole}$ by some few hundred meV. This shows that even for a wide band gap photo-catalyst like TiO$_2$ hole localization imposes quite some constraints on the efficiency of the reaction energetics. For efficient one-step water splitting using a single visible-light responsive photo-catalyst smaller band gap materials are required, typically with raised VBM positions like in the oxy-nitrides. One can expect even more severe constraints in these cases, underscoring the necessity to identify suitable co-catalysts also and in particular for the water oxidation half-reaction.

\begin{table}[ht]
	\caption{Calculated hole energies $\epsilon_{\rm hole}$ necessary to drive the water oxidation reaction downhill versus actual 
	surface hole trapping levels $\epsilon_{\rm T}$ present at different rutile and anatase TiO$_2$ facets (see text). All energies are 
	given with respect to vacuum. The surface hole trapping levels $\epsilon_{\rm T}$ are taken from ~\cite{Zawadzki12EES} and use 
	$\epsilon_{\rm VBM} = -7.1$\,eV ~\cite{Cheng10PRB} as reference.\label{tab:TiO2-hole}}
	\begin{ruledtabular}
	\begin{tabular}{ l c c}
		Facet           & $\epsilon_{\rm hole}$ [eV] & $\epsilon_{\rm T}$ [eV] \\ \hline
		Rutile  (110) & -7.5                       & -6.9                    \\
		Rutile  (001) & -7.3                       & -6.4                    \\ \hline
		Anatase (100) & -6.9                       & -6.6                    \\
		Anatase (001) & -7.0                       & -5.8
	\end{tabular}
	\end{ruledtabular}
\end{table}

\subsection{Small metal co-catalysts}

\begin{figure}[ht]
		\centering
		\includegraphics[scale=1.0,clip=true]{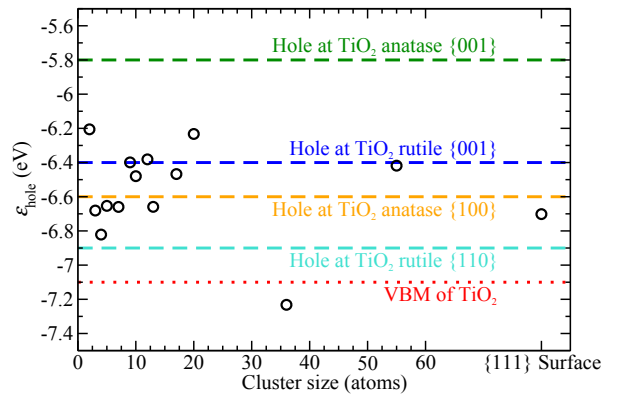}
		\caption{Computed hole energies $\epsilon_{\rm hole}$ necessary to drive the water oxidation half-reaction at Au clusters 
		in the size range between 2 and 55 atoms, as well as at (unreconstructed) Au(111). Additionally shown are the available
		hole trapping levels $\epsilon_{\rm T}$ at different TiO$_2$ facets, cf. Table \ref{tab:TiO2-hole}. All energies with respect to vacuum.}
		\label{fig:VBpred}
	\end{figure}

In this understanding we now proceed to apply our approach for the screening of co-catalysts. Particular efficiency emerges in this case when neglecting (for a first screening) any electronic and geometric changes of the co-catalyst particle due to adsorption at the photo-catalyst substrate, as well as any charge transfer between the two moieties. In this case, the thermodynamic theory allows to treat the reaction energetics at the co-catalyst separately. The discussion of suitable co-catalyst/photo-catalyst combinations only enters in a second step, when comparing the computed necessary $\epsilon_{\rm hole}$ at a given co-catalyst with available hole energies $\epsilon_{\rm T}$ at a given photo-catalyst. The actual DFT calculations for the co-catalyst are then simply performed at single-crystal surface models when aiming to assess the catalytic function of dominant facets of larger nanoparticles, or at clusters in the non-scalable size regime when aiming to assess potential nano-catalytic behavior. As a showcase we here present both avenues and for Au, which has shown much promise not only for its general nano-catalytic properties \cite{Valden1998SCI,Hakkinen2003ACIE,Yoon2005SCI,Vaida2010PSS}, but also in the water splitting context \cite{Subramanian2004JACS,Iwase2006CL,Berr2012NL}. Specifically, we compute the water oxidation pathway at (unreconstructed) Au(111) and at 13 clusters consisting of between $2$ and $55$ Au atoms. The obtained necessary hole energies $\epsilon_{\rm hole}$ are summarized in Fig. \ref{fig:VBpred} and compared to the afore discussed available hole energies $\epsilon_{\rm T}$ at different TiO$_2$ facets.

Intriguingly, already the inert Au(111) surface shows better water oxidation capabilities than most of the initially discussed TiO$_2$ facets, requiring a necessary hole energy that is with $\epsilon_{\rm hole} = -6.7$\,eV almost one eV smaller than the one required by rutile TiO$_2$(110). The reason for this lies in the much lower energy cost of step $A$ on Au, thus making step $C$ the new albeit much lower energetic bottleneck. The expected nano-catalytic properties are also clearly visible, with almost all tested clusters improving on the formation energy of adsorbed hydroperoxide (step $C$) and correspondingly requiring even smaller necessary hole energies. Particularly interesting are the small $\epsilon_{\rm hole}$ values obtained for Au$_{20}$ and Au$_{55}$, which could still be large enough particles to also exhibit suitable stability against sintering and corrosion. Comparing against the available hole energies also shown in Fig.~\ref{fig:VBpred} their $\epsilon_{\rm hole}$ values are still small enough to drive the reaction at most TiO$_2$ facets -- at least in the thermodynamic sense discussed here. In the spirit of computational screening, these two clusters would thus be appealing candidates for experimental testing, or for refining calculations explicitly addressing the kinetics along the reaction path. 

The microscopic reason for the comparatively low energy requirements (high values of $\epsilon_{\rm hole}$) of Au$_{20}$ and Au$_{55}$ lies in the adsorption energetics of the different reaction intermediates. While in TiO$_2$ the step determining $\epsilon_{\rm hole}$ -- i.e.~the most uphill step without driving force -- is the formation of adsorbed OH (step A, see Fig.~\ref{fig:dG-TiO2}), the picture is different for the gold clusters. For Au$_{20}$ the removal of the second proton and electron (step B) is determining but with $\Delta G_{B}=1.85$\,eV much less steeply uphill compared to TiO$_2$ and most other Au clusters (see Fig.~\ref{fig:VBpred}). In the case of Au$_{55}$ it is the formation of surface adsorbed OOH (step C) that is most uphill but again comparatively moderate with $\Delta G_{C}=2.04$\,eV. Thus, these two co-catalysts nicely demonstrate the fact that no clear size related trends can be expected from clusters in the non-scalable size regime.

Notwithstanding, with their $\epsilon_{\rm hole}$ values of -6.2\,eV and -6.4\,eV even Au$_{20}$ and Au$_{55}$, respectively, are still far away from an ideal co-catalyst function. Such a co-catalyst would exhibit a balanced free energy profile in which every reaction step along the sequence, cf. eqs.~\ref{eq:nr-orig}, goes uphill by 4.92\,eV/4 = 1.23\,eV, which would translate into a necessary hole energy of $\epsilon_{\rm hole} = - 5.6$\,eV. There is thus plenty of room left for improved co-catalyst materials and particle sizes, which can be efficiently screened with the approach presented here.

\section{Conclusions}
In conclusion we have presented a computational screening protocol to efficiently assess the suitability of co-catalyst/photo-catalyst combinations for the photo-oxidation of water. Our method, based on a thermodynamic approach popularised by N\o{}rskov and Rossmeisl, compares relative stabilities of reaction intermediates and thus allows prediction of hole energies necessary to drive the reaction at least thermodynamically downhill. The necessity for efficient co-catalysts in particular for the water oxidation half-reaction is emphasized by our hybrid-level DFT results, which indicate severe hole localization induced constraints on the reaction energetics even at wide band gap TiO$_2$ photo-catalysts. Results for Au co-catalysts show a trend towards smaller required driving forces, with improved catalytic behavior particularly observed for metal clusters in the non-scalable size regime. 

\section{Acknowledgements}
H.~O.~is supported by the Alexander von Humboldt foundation. The authors gratefully acknowledge support through the “Solar Technologies Go Hybrid” initiative of the State of Bavaria and the Gauss Centre for Supercomputing e.V.~(www.gauss-centre.eu) for funding this project by providing computing time on the GCS Supercomputer SuperMUC at Leibniz Supercomputing Centre (LRZ,  www.lrz.de).

\appendix
\section{Computational details}

All calculations were performed with the FHI-AIMS package \cite{Blum09CPC} with scaled ZORA relativistics, tight integration grids and a tier-2 numeric atomic orbital basis. Initial cluster geometries were taken from literature	 \cite{Doye1998NJC,Michaelian1999PRB,Hakkinen2000PRB,Assadollahzadeh2009JCP,Bao2009PRB} and re-optimised with the PBE \cite{Perdew1996PRL} generalised gradient (GGA) functional. For comparison with results by Vald{\'e}s and co-workers, \cite{Valdes08JPCC} TiO$_2$ surfaces were optimised on the GGA level with rPBE.\cite{Hammer1999PRB} For increased accuracy we then re-calculated total energies of all clusters and oxide surfaces with PBE0\cite{Adamo1999JCP}, while for the extended Au(111) surface we stayed at GGA level. Numerical convergence within $\pm 10$\,meV for $\epsilon_{\rm hole}$ was achieved at the chosen computational settings.

\subsection{Comparison of density functionals}
\begin{figure}[ht]
		\centering
		\includegraphics[scale=1.0,clip=true]{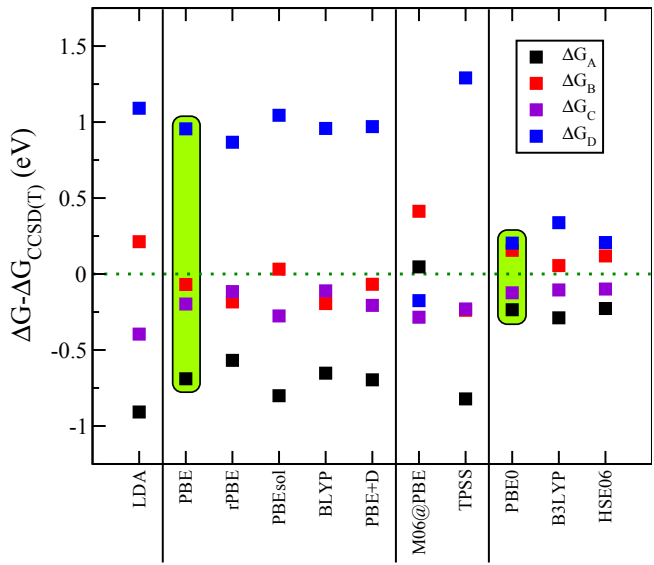}
		\caption{Free energies of steps A-D as predicted by different electronic structure methods (details in the text), compared to a CCSD(T) reference calculation. As a guide to the eye results for PBE and PBE0 have been marked in green.}
		\label{fig:Au2-dG-funcs}
\end{figure}

In order to find the optimal electronic structure method in terms of accuracy and computational effort, we first created for the Au$_2$ cluster references for free energies for all four steps with the coupled cluster single, double and perturbative triple substitutions (CCSD(T)) as implemented in the Gaussian09 package \cite{Gaussian09}. For these calculations we performed counterpoise corrections and extrapolated to the infinite basis set limit from double, triple, and quadruple zeta augmented, correlation consistent Dunning basis sets with polarised functions\cite{Kendall1992JCP,Peterson2005TCA}. We then proceeded to benchmark different DFT, hybrid-DFT and beyond-DFT methods against the CCSD(T) reference. The resulting deviations of the DFT-based methods from the CCSD(T) results are depicted in Fig.~\ref{fig:Au2-dG-funcs}. The benchmarked methods were separated according to Perdew's Jacob's ladder of density functionals: \cite{Perdew2001JACOB}
\begin{enumerate}
	\item LDA\cite{Perdew1992PRB}, the local density approximation
	\item Generalised gradient approximated (GGA) functionals: PBE \cite{Perdew1996PRL}, rPBE\cite{Hammer1999PRB}, PBEsol\cite{Perdew2008PRL}, BLYP\cite{Becke1988PRA,Lee1988PRB}, and PBE including Tkatchenko-Scheffler\cite{Tkatchenko2009PRL} dispersion correction (PBE+D)
	\item Meta-GGA's: M06\cite{Zhao2008TCA} as an energy post-correction on the basis of PBE orbitals (M06@PBE), and TPSS\cite{Zhao2008TCA}
	\item Hybrid functionals: PBE0\cite{Adamo1999JCP}, B3LYP\cite{Becke1993JCP}, and the range separated HSE06\cite{Krukau2006JCP}
\end{enumerate}
Benchmark calculations were performed with FHI-AIMS and tight, Tier 3 basis set settings. In order to eliminate geometry effects we conducted all benchmark calculations on the respective minimum energy configurations given by the PBE functional.

Note that the commonly used GGA level functionals show very large errors compared to the CCSD(T) references and there is only little improvement in going from PBE to rPBE. Of the meta-GGA's M06 fares already much better, while TPSS shows even worse results than LDA. The most consistent improvements are seen in the category of hybrid functionals with both PBE0 and HSE06 showing comparatively good results with a maximum deviation of $\approx200$ meV.

Extrapolating from our small test-system to other metal clusters in the non-scalable size regime, our benchmark calculations clearly show that LDA or GGA calculations can result in crass errors in the estimation of some free energy components. Specifically, step A the generation of an OH radical adsorbed on the cluster is severely underestimated in most functionals. While we expect a less dramatic picture for larger clusters--for which CCSD(T) would be prohibitively expensive--the conclusion to be drawn from these benchmarks is that in order to achieve predictive results we should use either PBE0 or HSE06.

\end{document}